\documentclass[pdflatex,sn-mathphys-num,iicol]{sn-jnl}% Math and Physical Sciences Numbered Reference Style 
\usepackage{graphicx}%
\usepackage{multirow}%
\usepackage{amsmath,amssymb,amsfonts}%
\usepackage{amsthm}%
\usepackage{mathrsfs}%
\usepackage[title]{appendix}%
\usepackage{xcolor}%
\usepackage{textcomp}%
\usepackage{manyfoot}%
\usepackage{booktabs}%
\usepackage{algorithm}%
\usepackage{algorithmicx}%
\usepackage{algpseudocode}%
\usepackage{listings}%
\usepackage{siunitx}
\usepackage{braket}
\newcommand{\Dbar}{\mathrm{\overline{D}}}
\newcommand{\dbar}{\mathrm{\overline{d}}}
\newcommand{\Hbar}{\mathrm{\overline{H}}}

\newcommand{\pbar}{\mathrm{\overline{p}}}

\newcommand{\electron}{\mathrm{e^-}}
\newcommand{\Ps}{\mathrm{Ps}}

\usepackage{tipa}
\newcommand{\lambdabar}{\mbox{\textipa{\textcrlambda}}}

\theoremstyle{thmstyleone}%
\theoremstyle{thmstyletwo}%

\theoremstyle{thmstylethree}%

\begin{document}

\title[Article Title]{Production and study of antideuterium with the GBAR beamline}

\author[1]{\fnm{Philipp} \sur{Blumer}}\email{philipp.blumer@cern.ch}
\author[2]{\fnm{Ben} \sur{Ohayon}}
\author[1]{\fnm{Paolo} \sur{Crivelli}}\email{paolo.crivelli@cern.ch}

\affil[1]{\orgdiv{Institute for Particle Physics and Astrophysics}, \orgname{\textsc{eth Zurich}}, \orgaddress{\city{Zurich}, \postcode{8093}, \country{Switzerland}}}
\affil[2]{\orgdiv{Physics Department}, \orgname{Technion—Israel Institute of Technology}, \orgaddress{Haifa}, \postcode{3200003}, \country{Israel}}


\abstract{The potential of circulating antideuterons ($\dbar$) in the AD/ELENA facility at CERN is under active investigation. Approximately 100 $\dbar$ per bunch could be delivered as a $\SI{100}{keV}$ beam based on measured cross-sections. These $\dbar$ could be further decelerated to $\SI{12}{keV}$ using the GBAR scheme, enabling the synthesis of antideuterium ($\Dbar$) via charge exchange with positronium, a technique successfully demonstrated with $\SI{6}{keV}$ antiprotons for antihydrogen production.
The AD/ELENA facility is currently studying the possibility of increasing the $\dbar$ rate using an optimized new target geometry. Assuming this is feasible, we propose further enhancing the anti-atom production by using laser-excited positronium in the $2P$ state within a cavity, which is expected to increase the $\Dbar(2S)$ production cross-section by almost an order of magnitude for $\dbar$ with $\SI{2}{keV}$ energy.
We present the projected precision for measuring the antideuterium Lamb shift and extracting the antideuteron charge radius, as a function of the beam flux.
}


\maketitle

\section{Introduction}\label{secIntro}

Antideuterium atoms ($\Dbar$), composed of an antideuteron ($\dbar$) and a positron, have yet to be experimentally observed. The successful synthesis and study of $\Dbar$  represents a frontier of antimatter research and would be a valuable tool in probing the longstanding puzzle of the matter-antimatter asymmetry~\cite{2003_DineKusenko,2012_Canetti}.

Significant achievements in antimatter physics have been made at CERN's Antiproton Decelerator (AD)~\cite{1996_AD}, further improved by the recent addition of the Extra Low Energy Antiproton (ELENA) ring~\cite{2016_ELENA,2022_ELENA}, by collaborations such as ATHENA, ATRAP, ALPHA, ASACUSA, BASE, AEGIS and GBAR~\cite{2002_ATHENA,2002_ATRAP,2016_ALPHA,2017_ALPHA,2018_ALPHA,2020_ALPHA,2021_ALPHA,2023_ALPHA,2011_ASACUSA,2013_ASACUSA,2021_AEGIS,2024_Aegis_PsLaser,2017_BASE,2022_BASE,2023_GBAR}. The various experiments at the Antimatter Facility provide exciting new possibilities for the potential production and study of molecular anti-ions~\cite{2018_Myers}, optical trapping of antihydrogen~\cite{2020_Crivelli} and the usage of a low energy $\dbar$ beam~\cite{2024_Caravita} for complementary tests of the Standard Model.

Antideuterons are expected to be produced at CERN's proton-to-antiproton converter target~\cite{1988_Johnson} and the potential of circulating $\dbar$ in the AD/ELENA facility is currently under consideration \cite{2024_ELENA,2024_ELENA_EXA}. 
We investigate the production of $\Dbar$ via charge exchange between $\dbar$ and positronium ($\Ps$) in the GBAR beamline:
\begin{equation}
\label{eqn:d-charge-exchange}
    \dbar + \Ps \rightarrow \Dbar + \electron.
\end{equation}
This same reaction is employed in GBAR to generate antihydrogen~\cite{2023_GBAR}, and extending it to $\Dbar$ could significantly advance our understanding of antimatter interactions at low energies.

\section{Antideuteron detection}

Antiprotons are produced by shooting \num{1.8e13} protons with an energy of $\SI{26}{GeV}$ from the CERN Proton Synchrotron (PS) onto an iridium target. Subsequently about \num{5e7} $\pbar$ with a momentum of $\SI{3.5}{GeV/c}$ are selected and guided to the Antiproton Decelerator (AD)~\cite{1996_AD}. 
The $\pbar$ are decelerated in several steps via stochastic and electron cooling to a kinetic energy of $\SI{5.3}{MeV}$. At this point, they are injected in the ELENA ring which further slows them down to $\SI{100}{keV}$ via electron cooling~\cite{2016_ELENA}. Four experiments can receive in parallel bunches of up to \num{1e7} $\pbar$ every $\SI{110}{\second}$~\cite{2024_ELENA}. The proton to antiproton conversion rate is of the order of $\SI{3e-6}{\pbar/\mathrm{p}}$, while the deceleration and transport efficiency to the experiments reaches up to $80\%$. 

The antideuteron production efficiency at the PS target is estimated to be around $\SI{4e-6}{\dbar/\pbar}$~\cite{1988_Johnson}. Therefore, one could expect to have a few hundred $\dbar$ decelerated and transported to the experiments. A dedicated study is ongoing to determine achievable $\dbar$ rates by increasing the incident proton energy, testing different target materials, and the subsequent cooling and transport to the experiments~\cite{2024_ELENA}. This work assumes that $\dbar$ are delivered with the same deceleration efficiency and energy distribution as the $\pbar$. This allows for no significant changes to the GBAR beamline, compared to the experimental setup used for $\Hbar$ production~\cite{2023_GBAR}.

Using an electrostatic drift tube, the pulsed anti-nucleons from the ELENA ring would be further decelerated from $\SI{100}{keV}$ to energies of less than $\SI{10}{keV}$~\cite{2021_GBAR_decelerator}. The $\dbar$ would then be focused using electrostatic lenses and annihilate on a Micro Channel Plate (MCP) detector at the end of the beamline, yielding a clear detection signal. The kinetic energy of the particles is given by the potential difference $\Delta V$ applied at the decelerator,
\begin{equation}
    E_{kin} = q\Delta V = \frac{m_\dbar v^2}{2},
\end{equation}
where $q$ is the charge and $v=\frac{x}{t}$ the velocity. By measuring the time of flight $t$ with the same configuration as in \cite{2023_GBAR}, the mass $m_\dbar$ could be determined to the percent level, i.e. four times more precise as the currently best-measured value $\SI{1867 \pm 80}{MeV/c^2}$~\cite{1965_Massam}.
For a stringent CPT and Lorentz invariance test, a precision measurement, such as the recent antiproton charge-to-mass-ratio by the BASE collaboration at the ppt level~\cite{2022_BASE}, would need to be considered.

\section{Antideuterium production}\label{secProd}

The $\Dbar$ production is given by eq.~\ref{eqn:d-charge-exchange} where $\dbar$ undergoes a charge exchange with $\Ps$. During an AD/ELENA cycle of about $\SI{2}{\minute}$, positrons are produced from an electron LINAC~\cite{2021_GBAR_positron}, accumulated in a buffer gas trap, and subsequently stacked in a high-field trap~\cite{2022_GBAR}. The positrons are extracted from the trap and implanted onto a mesoporous $\mathrm{SiO_2}$ film, that acts as a $\Ps$ converter at the intersection of the $\dbar$ beam axis. The triplet spin state, ortho-$\Ps$, has a lifetime of $\SI{142}{\nano\second}$ and diffuses out of the thin film, creating a cloud target for the charge exchange reaction.
The $\dbar$ would be focused through the $\Ps$ cloud, forming neutral $\Dbar$ traveling along the beamline. Charged particles are deflected using electrostatic fields before the $\Dbar$ would be detected on an MCP, see Fig.~\ref{fig:dbar_setup}. 

A Monte Carlo simulation, validated with $\Hbar$ data~\cite{2023_GBAR}, is adapted to this case.
It considers the time, energy, and position distributions of the $\Ps$ cloud and $\dbar$ beam, which were determined for the antihydrogen production with $\pbar$ at $\SI{6}{keV}$.
In the simulation, the initial $\Ps$ position is centered on the $\mathrm{SiO_2}$ film and according to the positron implantation profile represented with two Gaussian distributions of widths $\SI{3.5}{mm}$ and $\SI{0.5}{mm}$ in x and z direction respectively. The time evolution considers a positron pulse length of $\SI{17}{ns}$ (FWHM), a diffusion time of $\SI{10\pm 2}{ns}$ corresponding to an implantation energy of $\SI{4.3}{keV}$~\cite{2015_Deller}, and the $\Ps$ lifetime of $\SI{142}{ns}$. The atoms are modeled as being emitted with a \textit{cosine} distribution~\cite{2010_Cassidy} following a Maxwell-Boltzmann energy distribution at $\SI{750}{K}$. Although quantum mechanical effects from the confinement of $\Ps$ in the $\mathrm{SiO_2}$ pores suggest that the distribution is not thermal, the $\Ps$ velocities are well parameterized by it~\cite{2015_Deller,2020_Aegis, 2021_HeissPhD}. A cut-off on the minimal energy of the emitted $\Ps$, corresponding to the ground state energy in the pores is set to $\SI{45}{meV}$~\cite{2010_Crivelli}.

The antideuterons after the decelerator are expected to have a Gaussian profile with $\sigma_y=\SI{1.85}{mm}$ and $\sigma_z=\SI{2.78}{mm}$ in the y and z direction and a $\SI{100}{ns}$ bunch length, as was measured for $\pbar$~\cite{2023_GBAR}. They are modeled to have straight paths along the beam direction. 

The cross-section for neutral anti-atom charge exchange production has been calculated for $\Hbar$ with the Convergent Close Coupling (CCC) method~\cite{2016_Rawlins} and Coulomb-Born approximation (CBA)~\cite{2023_Hervieux}. The cross-section depends on the relative velocity between $\Ps$ and the nucleus and thus it can be assumed to be equal for the same velocities of $\pbar$ and $\dbar$. This assumption is supported by charge exchange measurements of protons and deuterons incident on a $\mathrm{Cs}$ vapor target~\cite{1975_Meyer}. Thus, for $\dbar$ with the same velocity as $\pbar$, the kinetic energy scales by a factor of 2, when $m_\dbar = 2\times m_\pbar$ as for ordinary matter. Finally, we set the $\dbar$ kinetic energy to $\SI{12}{keV}$, where the maximum cross-section is expected for $\Ps$ in the ground state~\cite{2023_GBAR,2016_Rawlins,2023_Hervieux}. The total ortho-$\Ps$ number per cycle is assumed to be $N_\Ps =\num{1e9}$. The $\Dbar$ production rate follows by mapping the $\Ps$ density at each time step with the $\dbar$ beam and multiplying with the cross-sections $\sigma_{CCC}=\SI{13.4e-16}{cm^2}$ or $\sigma_{CBA}=\SI{30.6e-16}{cm^2}$. 

\begin{figure}
    \centering
    \includegraphics[width=\linewidth]{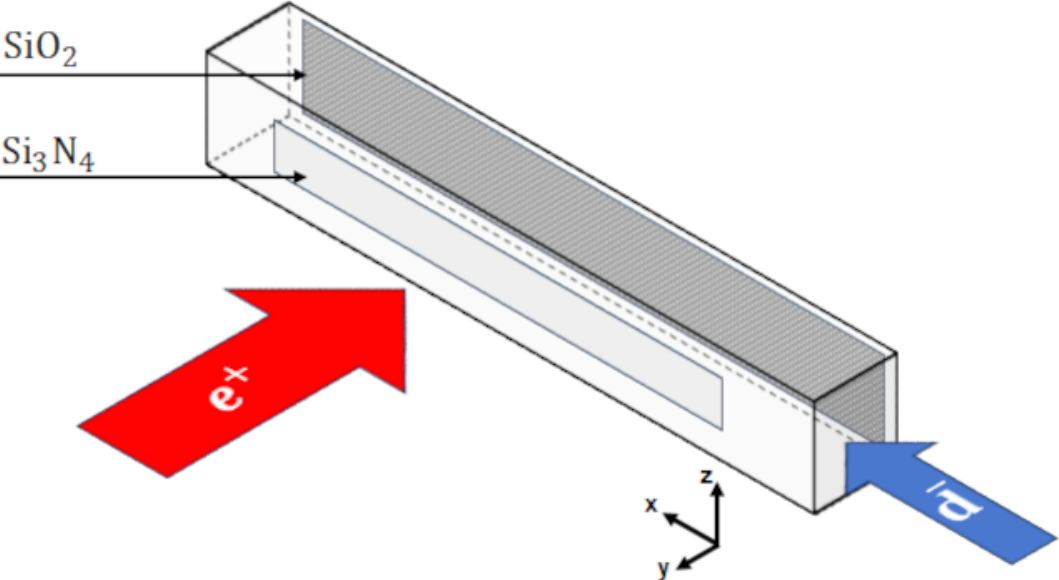}
    \caption{Schematic of the $\Dbar$ formation inside a cavity: Positrons are implanted into a porous $\mathrm{SiO_2}$ target and ortho-$\Ps$ diffuses. The $\dbar$ undergo a charge exchange with the $\Ps$ and form $\Dbar$. The neutral atoms continue traveling along a straight trajectory, separated from the charged nucleus using a static electric field.}
    \label{fig:cavity}
\end{figure}

Several upgrades to the beamline are outlined to increase the $\Dbar$ production yield. First, the flat $\Ps$ target would be replaced with a cavity of $\SI{2}{mm}\times\SI{2}{mm}\times\SI{20}{mm}$ coated on the inside with a $\mathrm{SiO_2}$ layer~\cite{2015_Cooke}. The positrons are extracted from the positron trap, shot through a $\mathrm{Si_3N_4}$ window of the cavity, and implanted onto the mesoporous $\mathrm{SiO_2}$ film, see Fig.~\ref{fig:cavity}. 
The resulting ortho-$\Ps$ is confined within the cavity and reflects off the walls due to the negative work function of the $\mathrm{SiO_2}$, which acts as a potential barrier~\cite{1998_Nagashima}. Experiments have shown comparable lifetimes of $\Ps$ in aerogels with $\SI{100}{nm}$ pores to its vacuum value~\cite{2007_Badertscher}, indicating that $\Ps$ is not sticking to the walls. Reflections on the rough internal surface are expected to be governed by a diffuse Lambert cosine law~\cite{1934_Knudsen} which is supported by advanced simulations~\cite{2008_Celestini} and has been experimentally demonstrated with muonium atoms~\cite{2016_Khaw}. These reflections increase the density of the $\Ps$ cloud, improving the production rate for $\Dbar$. 

Second, $\Ps$ inside the cavity would be laser-excited to the $2P$ state, having radiative and annihilation lifetimes of $\SI{3.2}{\nano\second}$ and $\SI{100}{\micro\second}$, respectively. As demonstrated in Ref.~\cite{2024_Aegis_PsLaser,2024_Shu}, the $1^3S-2^3P$ transition in $\Ps$ can be effectively saturated using a $\SI{243}{nm}$ UV laser. Measurements of $\Ps(2P)$ excitation inside of porous silica~\cite{2011_Cassidy} and the comparable lifetimes observed within aerogels and vacuum suggest that collisions with the silica film on the cavity walls are elastic, with negligible quenching to the ground state. Within the saturation regime, the annihilation rate for $\Ps$ can be expressed as: 
\begin{equation}
    \gamma_\mathrm{saturated} = \frac{\gamma_{1S}+\gamma_{2P}}{2} \approx \frac{1}{\SI{284}{ns}},
\end{equation}
reflecting the increased lifetime.
Additionally, for excited $\Ps(2P)$, the charge exchange cross-section increases drastically, by a factor $\approx 6 - 160$, when the $\dbar$ kinetic energy is $\SI{2}{keV}$~\cite{2016_Rawlins,2023_Hervieux}.

A Penning-Malmberg trap was installed after the electrostatic decelerator in GBAR, to reduce the antinucleon energy spread via electron cooling. We assume the simulation results for $\pbar$~\cite{2022_Yoo} are also valid for $\dbar$ and that they could be extracted with $\SI{2}{keV}$ energy from the trap, a reduced beam width of $\sigma_y=\SI{0.677}{mm}$ and $\sigma_z=\SI{0.512}{mm}$, and with a shortened bunch length of $\SI{56}{ns}$. The trap allows further stacking of the $\dbar$ to increase the number of anti-nucleons per bunch and the process of recycling anti-nucleons in the GBAR beamline~\cite{2018_Husson}. 

\begin{figure}[t!]
    \centering
    \includegraphics[width=\linewidth]{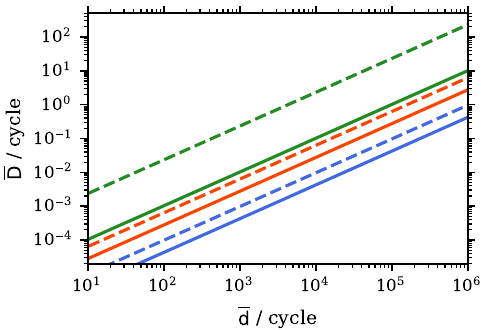}
    \caption{$\Dbar$ production rate assuming $10^9$ ortho-$\Ps$ from a flat target interact with antideuterons with $\SI{12}{keV}$ (blue) or inside a cavity with $\SI{2}{keV}$ (orange). Inside the cavity, $\Ps$ can be excited to the $2P$ state, further increasing the charge exchange cross-section (green) calculated with the Convergent Close Coupling (CCC, solid) method~\cite{2016_Rawlins} and Coulomb-Born approximation (CBA, dashed)~\cite{2023_Hervieux}.}
    \label{fig:dbar_prod}
\end{figure}

For spectroscopy measurements of $\Dbar$ it is important to produce as many anti-atoms as possible. Achieving higher rates of $\dbar$ requires a dedicated study, including at least a new target geometry to maximize the $\dbar$ production and adapted stochastic- and electron-cooling~\cite{2024_ELENA,2024_ELENA_EXA}. We present in Fig.~\ref{fig:dbar_prod} the linear dependence of the $\Dbar$ rate for potentially increasing $\dbar$ per ELENA cycle. Using a cavity and laser excited $\Ps$, a rate in the order of $0.1\,\Dbar/\text{cycle}$ would be feasible with $10^4\,\dbar/\text{cycle}$. 
Assuming the same behavior as for antihydrogen, it is expected that in the case of ground state $\Ps$ and $\Ps(2P)$, the $\Dbar(2S)$ population is between $10-20\%$~\cite{2016_Rawlins,2023_Hervieux}.

\section{Antideuterium Lamb shift}\label{secLS}
The measurement of the $2S_{1/2} \rightarrow 2P_{1/2}$ energy transition, known as the Lamb shift, was accomplished for the first time in atomic hydrogen in 1947~\cite{1947_Lamb,1956_Lamb}. This energy splitting is not predicted by the Dirac theory~\cite{1928_Dirac}, which considers the principles of quantum mechanics and special relativity. The observation of the non-degenerate $2S_{1/2}$ and $2P_{1/2}$ states led to the development of quantum electrodynamics (QED)~\cite{1946_Tomonaga}. 

The largest contributions to the $\Dbar$ Lamb shift are the QED effects of electron self-energy ($\sim\SI{1}{GHz}$) and vacuum polarization ($\sim\SI{-27}{MHz}$) from virtual electron-positron pairs. Of special interest is the nuclear size effect: 
\begin{equation}
    E_{nucl} = \frac{2}{3}m_e c^2 \frac{(Z\alpha)^4}{n^3}\left(\frac{m_r}{m_e}\right)^3\left(\frac{r_\dbar}{\lambdabar_C}\right)^2 \delta_{l0},
\end{equation}
where $m_e$ is the electron mass, $m_r=\frac{m_e m_\dbar}{m_e+m_\dbar}$ the reduced mass with the mass of the nucleus $m_\dbar$, $c$ is the speed of light in vacuum, $\alpha$ the fine-structure constant, $Z$ the atomic number, $r_\dbar$ the root-mean-squared (RMS) charge radius of the nucleus, $\lambdabar_C=\hbar/(m_e c)$ is the reduced Compton wavelength, $n$ the principle quantum number and $\delta_{l0}$ is the Kronecker delta depending on the orbital angular momentum quantum number $l$~\cite{2018_CODATA}. In contrast to (anti-)hydrogen, $E_{nucl}$ is 6.5 times larger for (anti-)deuterium due to the 2.5 times larger (anti-)deuteron radius~\cite{2018_CODATA}. Thus, measuring the Lamb shift to the precision level of $\SI{1}{MHz}$ the $\dbar$ RMS charge radius could be determined for the first time at the $10\%$ level.

\begin{figure}[t!]
    \centering
    \includegraphics[width=\linewidth]{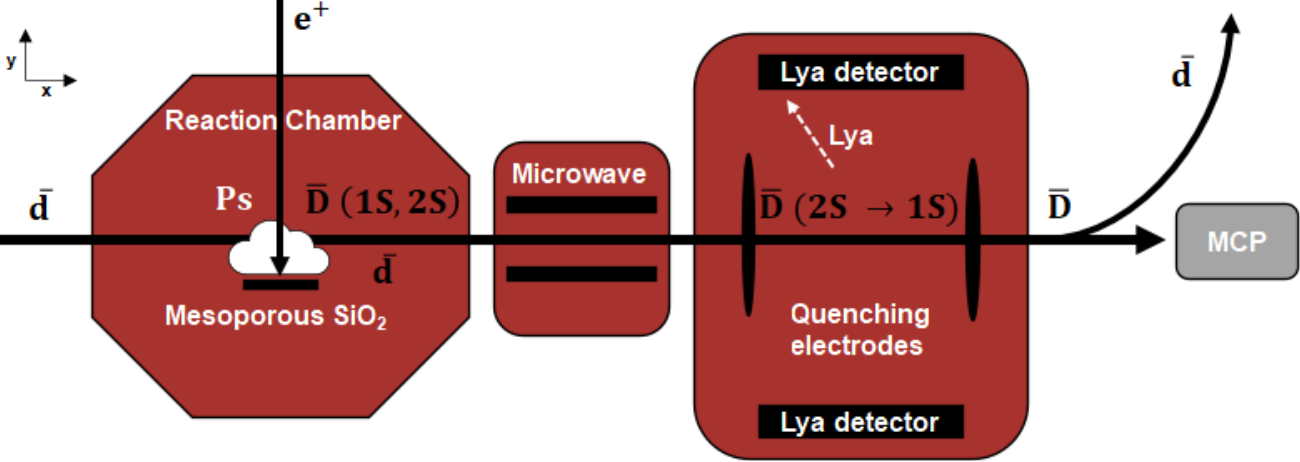}
    \caption{Inside the reaction chamber $\dbar$ undergo a charge-exchange reaction with $\Ps$ inside a cavity coated with $\mathrm{SiO_2}$, synthesizing $\Dbar$ with a fraction being in the $2S$ state. The anti-atoms traverse a microwave field oscillating at a frequency $\omega$, which induces the transitions to the short-lived $2P$ states. These $2P$ states deexcite within $\tau=\SI{1.6}{\nano\second}$ to the ground state by emitting Lyman-$\alpha$ photons with a wavelength of $\SI{121}{\nano\meter}$. The surviving $\Dbar(2S)$ population is quenched to the $2P$ state using a static electric field, and the emitted Lyman-$\alpha$ photons are detected with specially coated MCP detectors. At the end of the beamline, charged particles are deflected by a static electric field, ensuring that only neutral $\Dbar$ atoms are detected on the final MCP. By analyzing the coincidence signals of the Lyman-$\alpha$ photons and the stopping detector, the Lamb shift is determined by measuring the $2S$ population as a function of the microwave frequency $\omega$.}
    \label{fig:dbar_setup}
\end{figure}

\begin{figure}[t]
    \centering
    \includegraphics[width=\linewidth]{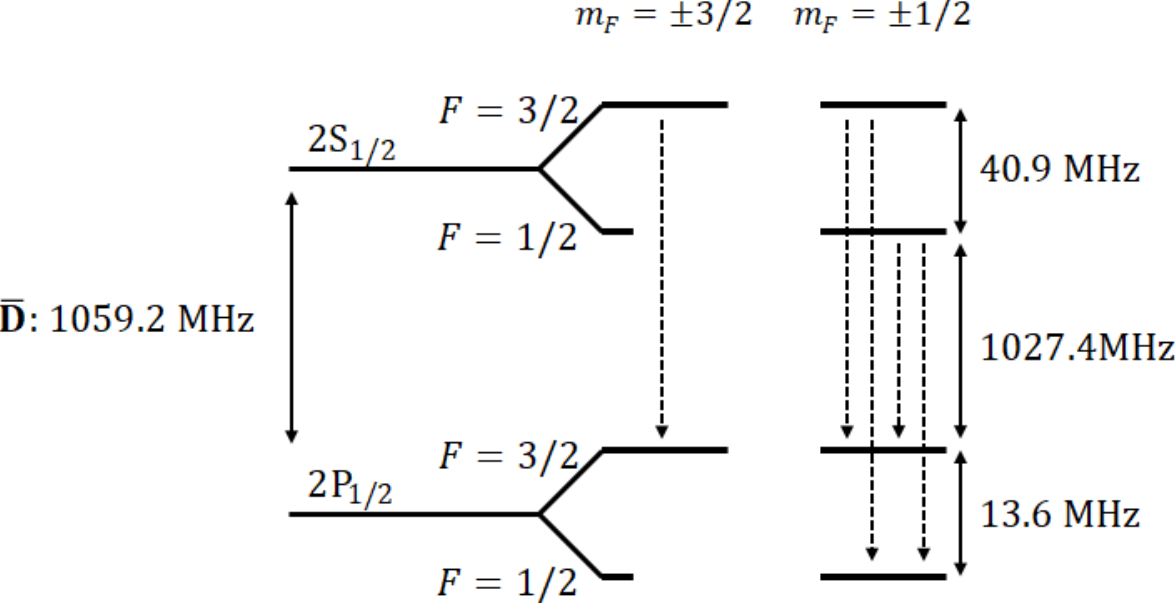}
    \caption{Sketch of the $2S_{1/2}$ and $ 2P_{1/2}$ hyperfine states for $\Dbar$. The dotted arrows represent the allowed energy transitions.}
    \label{fig:e_level}
\end{figure}

The Lamb shift experiment for $\Dbar$ is adapted from the proposal for $\Hbar$, currently installed at the GBAR beamline~\cite{2016_Crivelli} and is sketched in Fig.~\ref{fig:dbar_setup}. With a similar apparatus, the MuMASS collaboration measured the Lamb shift of muonium to a precision of $\SI{2.5}{MHz}$~\cite{2022_Ohayon,2022_Janka} at the Paul Scherrer Institute (PSI) in Switzerland. As shown in Sec. \ref{secProd}, at least $10\%$ of the atoms are expected to be produced in the metastable $2S$ state before they relax to the ground state. The $\Dbar(2S)$ pass a microwave field region oscillating at a frequency $\omega$ where the transition to the $2P_{1/2}$ state is induced. 
Fig.~\ref{fig:e_level} shows the energy levels of the $2S_{1/2}$ and $2P_{1/2}$ hyperfine states in $\Dbar$ and their allowed electric dipole transitions. Anti-atoms in the $2P$ state have a short lifetime of $\tau_{2P}=\SI{1.6}{ns}$ after which they deexcite via emission of a $\SI{121}{nm}$ Lyman-$\alpha$ photon to the ground state. The natural linewidth $\Delta \nu$ of the frequency transitions arises due to the Heisenberg uncertainty principle, relating the lifetime of the excited state to its energy spread. Experimentally, this results in a frequency spread of the order $\Delta\nu = 1/(2\pi\tau_{2P}) \approx \SI{100}{MHz}$, even under ideal conditions. 
After the microwave region, the remaining $2S$ atoms reach the detection chamber, where they are quenched to the $2P$ state using a static electric field of the order of $\SI{250}{V/cm}$. The emitted Lyman-$\alpha$ photons are detected with $\mathrm{CsI}$-coated MCP detectors, and the total photon detection efficiency of the setup is estimated to be $\epsilon=16\%$~\cite{2022_JankaPhD}.
At the end of the beamline, another static electric field deflects charged particles such that only neutral anti-atoms are detected on the final MCP. Analyzing the coincidence signal between Lyman-$\alpha$ photons and $\Dbar$ on the specific MCPs allows us to measure the $2S$ population as a function of the microwave frequency $\omega$.

Determining the resonance center $\nu$ has a limited precision $\delta \nu$ that depends on the number of detected Lyman-$\alpha$ photons $N$ and the transition linewidth $\Delta \nu$. A rule of thumb for estimating $\delta \nu$ is derived from signal-to-noise ratio considerations in spectroscopy, reflecting statistical limitations under Gaussian noise. Assuming that the linewidth dominates the uncertainty, $\delta \nu$ is proportional to $\Delta \nu$, as broader linewidths lead to greater uncertainty in determining the line center. However, the precision improves with the square root of the number of detected events, as more data points refine the line position. The relationship
\begin{equation}
    \delta\nu = \Delta\nu/\sqrt{N}
\end{equation}
has been shown to work well for the muonium measurements~\cite{2022_Ohayon,2022_Janka}.

Assuming that an upgraded AD/ELENA facility produces $\num{1e4}\,\dbar/\text{cycle}$ and with that $0.01\,\Dbar/\text{cycle}$ in the $2S_{1/2}$ state, it would take about 90 days for a first $\SI{10}{MHz}$ Lamb shift measurement.

\section{Conclusion}\label{secConc}
We have shown that with the currently estimated $\dbar$ rates produced at the AD/ELENA facility, the detection of $\Dbar$ and measurement of $m_\dbar$ are promising at the existing GBAR beamline. 
To enhance $\Dbar$ production, we propose exciting $\Ps$ inside a cavity coated with $\mathrm{SiO_2}$ to the $2P$ state using a $\SI{243}{nm}$ laser. This significantly increases the $\Dbar$ production cross-section.
Based on a Monte Carlo simulation, we estimate that $\num{1e4}\,\dbar/\text{cycle}$ would enable a measurement of the $\Dbar$ Lamb shift with a precision of $\SI{10}{MHz}$ within 90 days. A tenfold improvement in precision would allow the determination of the $\dbar$ charge radius at the $10\%$ level, providing a complementary test of CPT symmetry. However, such a $\dbar$ rate would require significant upgrades to the current facility, which is currently being investigated by the AD/ELENA team.
\backmatter

\subsection*{Acknowledgments}
The Swiss National Science Foundation has supported this work under grants 197346 and 201465.

\subsection*{Author contributions}
All authors contributed to the study's conception and design. PB performed simulation and analysis. PB wrote the first draft of the manuscript.
All authors commented on previous versions of the manuscript. All authors read and approved the final manuscript.

\subsection*{Data Availability Statement}
 The datasets generated during and/or analyzed during the current study are available from the corresponding author.

\bibliography{bibliography}

\end{document}